\documentclass[11pt]{article}
\usepackage{a4wide}
\usepackage{amsmath}
\usepackage[dvips]{graphics,color}
\usepackage{epsfig}
\usepackage{amssymb}
\usepackage{amsthm}
\usepackage{changebar}
\def\ci{\perp\kern-.5em\perp}

\def\cov{\text{Cov}}

\begin{document}

\title{A note on state space representations of locally stationary wavelet time series}

\author{K. Triantafyllopoulos\footnote{Department of
Probability and Statistics, Hicks Building, University of Sheffield,
Sheffield S3 7RH, UK, email: {\tt
k.triantafyllopoulos@sheffield.ac.uk}} \and G.P.
Nason\footnote{Department of Mathematics, University of Bristol,
Bristol, UK}}

\date{\today}

\maketitle

\begin{abstract}

In this note we show that the locally stationary wavelet process can
be decomposed into a sum of signals, each of which following a
moving average process with time-varying parameters. We then show
that such moving average processes are equivalent to state space
models with stochastic design components. Using a simple simulation
step, we propose a heuristic method of estimating the above state
space models and then we apply the methodology to foreign exchange
rates data.

\textit{Some key words:} wavelets, Haar, locally stationary process,
time series, state space, Kalman filter.
\end{abstract}

\section{Introduction}\label{wavelet}

\begin{figure}[t]
\begin{center}
 \epsfig{file=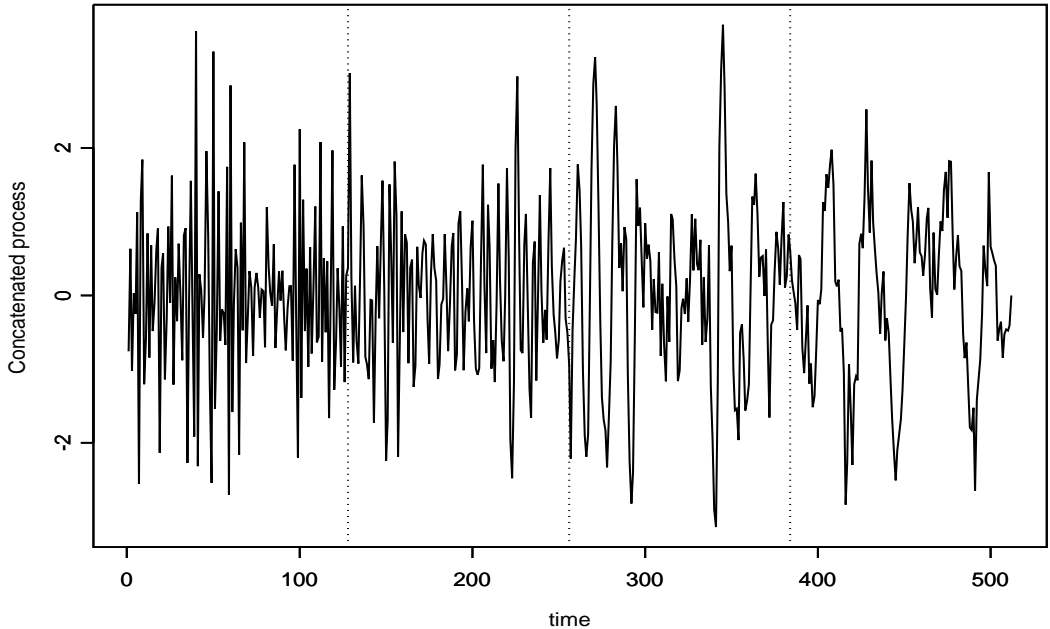, height=10cm, width=15cm}
 \caption{Concatenation of four MA time series with different
    parameters. Overall the process is not stationary.
    The dotted vertical lines indicate the transition between one
    MA process and the next.\label{fig1}}
 \end{center}
\end{figure}

Nason {\it et al.} (2000) define a class of locally stationary time
series making use of non-decimated wavelets. Let $\{y_t\}$ be a
scalar time series, which is assumed to be locally stationary, or
stationary over ceratin intervals of time (regimes), but overall
non-stationary. For more details on local stationarity the reader is
referred to Dahlhaus (1997), Nason {\it et el.} (2000), Francq and
Zakoan (2001), and Mercurio and Spokoiny (2004). For example, Figure
\ref{fig1} shows the nonstationary process considered in Nason {\it
et al.} (2000), which is the concatenation of 4 stationary moving
average processes, but each with different parameters. We can see
that within each of the four regimes, the process is weakly
stationary, but overall the process is non-stationary.

The locally stationary wavelet (LSW) process is a doubly indexed
stochastic process, defined by
\begin{equation}\label{lsw}
y_{t}=\sum_{j=-J}^{-1}\sum_{k=0}^{T-1}w_{jk}\psi_{j,t-k}\xi_{jk},
\end{equation}
where $\xi_{jk}$ is a random orthonormal increment sequence (below
this will be iid Gaussian) and $\{\psi_{jk}\}_{j,k}$ is a discrete
non-decimated family of wavelets for $j=-1,-2,\ldots,-J$,
$k=0,\ldots,T-1$, based on a mother wavelet $\psi (t)$ of compact
support. Denote with $I_A(x)$ the indicator function, i.e.
$I_A(x)=1$, if $x\in A$ and $I_A=0$, otherwise. The simplest class
of wavelets are the Haar wavelets, defined by
$$
\psi_{jk}=2^{j/2} I_{\{0,\ldots,2^{-j-1}-1\}}(k)
-2^{j/2}I_{\{2^{-j-1},\ldots,2^{-j}-1\}}(k),
$$
for $j\in \{-1,-2,\ldots,-J\}$ and $k\in
\{\ldots,-2,-1,0,1,2,\ldots\}$, where $j=-1$ is the finest scale. It
is also assumed that $E(\xi_{jk})=0$, for all $j$ and $k$ and so
$y_{t}$ has zero mean. The orthonormality assumption of
$\{\xi_{jk}\}$ implies that $\cov(\xi_{jk},\xi_{\ell
m})=\delta_{j\ell}\delta_{km}$, where $\delta_{jk}$ denotes the
Kronecker delta, i.e. $\delta_{jj}=1$ and $\delta_{jk}=0$, for
$j\neq k$.

The parameters $w_{jk}$ are the amplitudes of the  LSW process. The
quantity $w_{jk}$ characterizes the amount of each oscillation,
$\psi_{j,t-k}$ at each scale, $j$, and location, $k$ (modified by
the random amplitude, $\xi_{jk}$). For example, a large value of
$w_{jk}$ indicates that there is a chance (depending on $\xi_{jk}$)
of an oscillation, $\psi_{j, t-k}$, at time $t$. Nason {\em et al.}
(2000) control the
 evolution of the statistical characteristics of $y_{t}$ by coupling $w_{jk}$ to a function
 $W_j (z)$ for $z \in (0,1)$ by
 $w_{jk } = W_j (k/T) + {\mathcal O} (T^{-1})$.
Then, the smoothness properties of $W_j (z)$ control the possible
rate of change of $w_{jk}$ as a function of $k$, which consequently
controls the evolution of the statistical properties of $y_{t}$. The
smoother $W_j(z)$ is, as a function of $z$, the slower that $y_{t}$
can evolve. Ultimately, if $W_j(z)$ is a constant function of $z$,
then $y_{t}$ is weakly stationary.

The non-stationarity in the above studies is better understood as
local-stationarity so that the $w_{jk}$'s are close to each other.
To elaborate on this, if $w_{jk}=w_j$ (time invariant), then $y_{t}$
would be weakly stationary. The attractiveness of the LSW process,
is its ability to consider time-changing $w_{jk}$'s.

Nason {\it et al.} (2000) define the evolutionary wavelet spectrum
(EWS) to be $S_j (z) = | W_j (z) |^2$ and discuss methods of
estimation. Fryzlewicz {\it et al.} (2003) and Fryzlewicz (2005)
modify the LSW process to forecast log-returns of non-stationary
time series. These authors analyze daily FTSE 100 time series using
the LSW toolbox. Fryzlewicz and Nason (2006) estimate the EWS by
using a fast Haar-Fisz algorithm. Van Bellegem and von~Sachs (2008)
consider adaptive estimation for the EWS and permit jump
discontinuities in the spectrum.

In this paper we show that the process $y_{t}$ can be decomposed
into a sum of signals, each of which follows a moving average
process with time-varying parameters. We deploy a heuristic approach
for the estimation of the above moving average process and an
example, consisting of foreign exchange rates, illustrates the
proposed methodology.

\section{Decomposition at scale $j$}\label{scalej}

The LSW process (\ref{lsw}) can be written as
\begin{equation}\label{s3eq1}
y_{t}=\sum_{j=-J}^{-1}x_{jt},
\end{equation}
where
\begin{equation}\label{s3eq2}
x_{jt}=\sum_{k=0}^{T-1}w_{jk}\psi_{j,t-k}\xi_{jk}.
\end{equation}
For computational simplicity and without loss in generality, we omit
the minus sign of the scales $(-J,\ldots,-1)$ so that the summation
in equation (\ref{s3eq1}) is done from $j=1$ (scale $-1$) until
$j=J$ (scale $-J$).

Using Haar wavelets, we can see that at scale 1, we have from
(\ref{s3eq2}) that $x_{1t}=\psi_{1,0}w_{1t}\xi_{1t}
+\psi_{1,-1}w_{1,t-1}\xi_{1,t-1}$, since there are only 2 non-zero
wavelet coefficients. Then we can re-write (\ref{s3eq2}) as
$x_{1t}=\alpha_{1t}^{(0)}\xi_{1t}+\alpha_{1t}^{(1)}\xi_{1,t-1}$,
which is a moving average process of order one, with time-varying
parameters $\alpha_{1t}^{(0)}$ and $\alpha_{1t}^{(1)}$. This process
can be referred to as TVMA$(1)$ process.

In a similar way, for any scale $j=1,\ldots,J$, we can write
$$
x_{j t}
=\psi_{j,0}w_{jt}\xi_{jt}+\psi_{j,-1}w_{j,t-1}\xi_{j,t-1}+\cdots
+\psi_{j,-2^j+1}w_{j,t-2^j+1}\xi_{j,t-2^j+1}
$$
so that we obtain the TVMA($2^j-1$) process
\begin{equation}\label{tvma:j}
x_{jt}=\alpha_{jt}^{(0)}\xi_{jt}+a_{jt}^{(1)}\xi_{j,t-1}+\cdots
+a_{jt}^{(2^j-1)}\xi_{j,t-2^j+1},
\end{equation}
where $\alpha_{jt}^{(\ell)}=\psi_{j,-\ell}w_{j,t-\ell},$ for all
$\ell=0,1,\ldots ,2^j-1$ and $j=1,\ldots,J$. Thus the process
$y_{t}$ is the sum of $J$ TVMA processes. However, we note that not
all time-varying parameters $a_{jt}^{(\ell)}$
$(\ell=0,1,\ldots,2^j-1)$ are independent, since, for a fixed $j$,
they are all functions of the $\{w_{jt}\}$ series.

We advocate that $w_{jt}$ is a signal and as such we treat it as an
unobserved stochastic process. Indeed, from the slow evolution of
$w_{jt}$, we can postulate that $w_{jt}-w_{j,t-1}\approx 0$, which
motivates a random walk evolution for $w_{jt}$ or
$w_{jt}=w_{j,t-1}+\zeta_{jt}$, where $\zeta_{jt}$ is a Gaussian
white noise, i.e. $\zeta_{jt}\sim N(0,\sigma_j^2)$, for $\sigma^2_j$
a known variance, and $\zeta_{jt}$ is independent of $\zeta_{kt}$,
for all $j\neq k$. The magnitude of the differences between $w_{j,
t-1}$ and $w_{j t}$ can be controlled by $\sigma^2_j$ and this
controls on the degree of evolution of $w_{jt}$ as a function of $t$
and hence on $y_t$ through (\ref{s3eq1}).

At scale 1 we can write $x_{1t}$ as
$$
x_{1t}=\psi_{1,0}w_{1t}\xi_{1t}+\psi_{1,-1}w_{1,t-1}\xi_{1,t-1} =
(\psi_{1,0}\xi_{1t}+\psi_{1,-1}\xi_{1,t-1})w_{1,t-1}+\psi_{1,0}\zeta_{1t}\xi_{1t},
$$
where we have used $w_{1t}=w_{1,t-1}+\zeta_{1t}$. Likewise at scale
2 we have
\begin{eqnarray*}
x_{2t}&=& \psi_{2,0}w_{2t}\xi_{2t}+\psi_{2,-1}w_{2,t-1}\xi_{2,t-1}+
\psi_{2,-2}w_{2,t-2}\xi_{2,t-2}+\psi_{2,-3}w_{2,t-3}\xi_{2,t-3} \\
&=&
(\psi_{2,0}\xi_{2t}+\psi_{2,-1}\xi_{2,t-1}+\psi_{2,-2}\xi_{2,t-2}+\psi_{2,-3}\xi_{2,t-3})w_{2,t-3}
\\
&&+\psi_{2,0}\zeta_{2,t-2}\xi_{2t}+\psi_{2,0}\zeta_{2,t-1}\xi_{2t}+\psi_{2,0}\zeta_{2t}\xi_{2t}
\\ && +
\psi_{2,-1}\zeta_{2,t-2}\xi_{2,t-1}+\psi_{2,-1}\zeta_{2,t-1}\xi_{2,t-1}
\\ &&+\psi_{2,-2}\zeta_{2,t-2}\xi_{2,t-2},
\end{eqnarray*}
where we have used $w_{2,t-2}=w_{2,t-3}+\zeta_{2,t-2}$,
$w_{2,t-1}=w_{2,t-3}+\zeta_{2,t-2}+\zeta_{2,t-1}$ and
$w_{2t}=w_{2,t-3}+\zeta_{2,t-2}+\zeta_{2,t-1}+\zeta_{2t}$.

In general we observe that at any scale $j=1,\ldots,J$ we can write
\begin{equation}\label{eq:dec1}
x_{jt}=\sum_{k=0}^{2^j-1}\psi_{j,-k}\xi_{j,t-k}w_{j,t-2^j+1} +
\sum_{k=0}^{2^j-2}\sum_{m=k}^{2^j-2}\psi_{j,-k}\xi_{j,t-k}\zeta_{j,t-m},
\quad t=2^j,2^j+1,\ldots,
\end{equation}
where the $w_{jt}$'s follow the random walk
\begin{equation}\label{eq:dec2}
w_{j,t-2^j+1}=w_{j,t-2^j}+\zeta_{j,t-2^j+1}, \quad
\zeta_{j,t-2^j+1}\sim N(0,\sigma^2_j).
\end{equation}

\section{A state space representation}\label{tvma}

For estimation purposes one could use a time-varying moving average
model in order to estimate $\{w_{jk}\}$ in (\ref{tvma:j}). Moving
average processes with time-varying parameters are useful models for
locally stationary time series data, but their estimation is more
difficult that that of time-varying autoregressive processes
(Hallin, 1986; Dahlhaus, 1997). The reason for this is that the
time-dependence of the moving average coefficients may result in
identifiability problems. The consensus is that some restrictions of
the parameter space of the time-varying coefficients should be
applied; for more details the reader is referred to the above
references as well as to Triantafyllopoulos and Nason (2007).

In this section we use a heuristic approach for the estimation of
the above models. First we recast model
(\ref{eq:dec1})-(\ref{eq:dec2}) into state space form. To end this
we write
\begin{equation}\label{ss1}
x_{jt}=A_{jt}w_{j,t-2^j+1}+\nu_{jt},
\end{equation}
where $A_{jt}=\sum_{k=0}^{2^j-1}\psi_{j,-k}\xi_{j,t-k}$ and
$\nu_{jt}=\sum_{k=0}^{2^j-2}\sum_{m=k}^{2^j-2}\psi_{j,-k}\xi_{j,t-k}\zeta_{j,t-m}$,
for $t=2^j,2^j+1,\ldots$. In addition we assume that $\xi_{jt}^i$ is
independent of $\zeta_{js}^i$, for $i=1,2$ and for any $t,s$, so
that
\begin{equation}\label{obserror}
\nu_{jt}\sim
N\left\{0,\sigma^2_j\sum_{k=0}^{2^j-1}\psi_{j,-k}^2(2^j-k-1)\right\}.
\end{equation}
Equations (\ref{ss1}), (\ref{eq:dec2}), (\ref{obserror}) define a
state space model for $x_{jt}$ and by defining
$A_t=(A_{1t},\ldots,A_{Jt})'$ and by noting that $\nu_{jt}$ is
independent of $\nu_{kt}$, for any $j\neq k$, we obtain by
(\ref{s3eq1}) a state space model for $y_t$, which essentially is
the superposition of $J$ state space models of the form of
(\ref{ss1}), (\ref{eq:dec2}), (\ref{obserror}), each being a state
space model for each scale $j=1,\ldots,J$.

Given a set of data $y^T=\{y_1,\ldots,y_T\}$, a heuristic way to
estimate $\{w_{jt}\}$, is to simulate independently all $\xi_{jt}$
from $N(0,1)$, thus to obtain simulated values for $A_{jt}$ and
then, conditional on $A_t$, to apply the Kalman filter to the state
space model for $y_t$. This procedure will give simulations from the
posterior distributions of $w_{jt}$ and also from the predictive
distributions of $y_{t+h}|y^t$. The estimator of $w_{jt}$ and the
forecast of $y_{t+h}$ are conditional on the simulated values of
$\{\xi_{jt}\}$. For competing simulated sequences $\{\xi_{jt}\}$ the
performance of the above estimators/forecasts can be judged by
comparing the respective likelihood functions (which are easily
calculable by the Kalman filter) or by comparing the respective
posterior and forecast densities (by using sequential Bayes
factors). Another means of model performance may be the computation
of the mean square forecast error.

\begin{figure}[t]
\begin{center}
 \epsfig{file=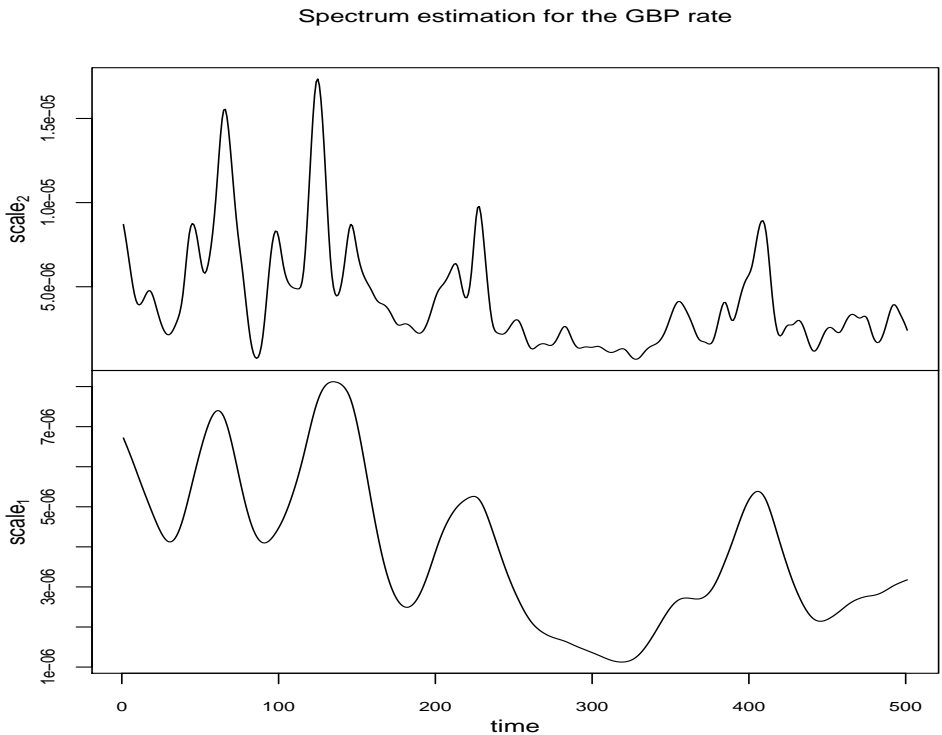, height=10cm, width=15cm}
 \caption{Simulated values of posterior estimates of $\{S_{jt}=w_{jt}^2\}$, for $\{y_{1t}\}$ (GBP rate). Shown are
 simulations of $\{S_{1t}\}$ and $\{S_{2t}\}$,
 corresponding to scales 1 and 2.}\label{fig2}
 \end{center}
\end{figure}

\begin{figure}[t]
\begin{center}
 \epsfig{file=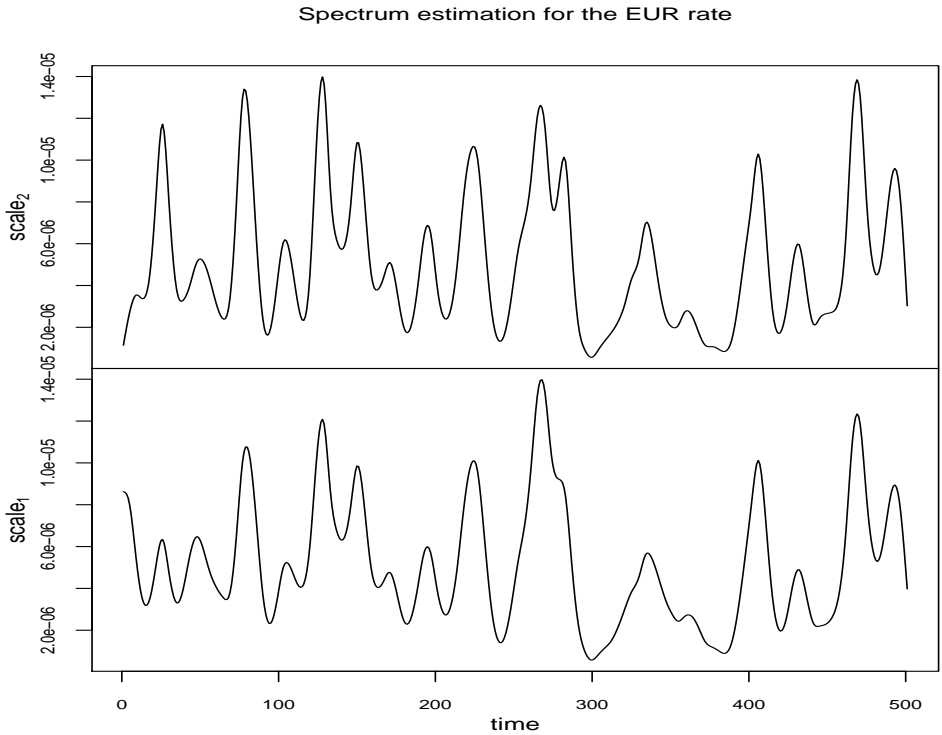, height=10cm, width=15cm}
 \caption{Simulated values of posterior estimates of $\{S_{jt}=w_{jt}^2\}$, for $\{y_{2t}\}$ (EUR rate). Shown are
 simulations of $\{S_{1t}\}$ and $\{S_{2t}\}$,
 corresponding to scales 1 and 2.}\label{fig3}
 \end{center}
\end{figure}

We illustrate this approach by considering foreign exchange rates
data. The data are collected in daily frequency from 3 January 2006
to and including 31 December 2007 (considering trading days there
are 501 observations). We consider two exchange rates: US dollar
with British pound (GBP rate) and US dollar with Euro (EUR rate).
After we transform the data to the log scale, we propose to use the
LSW process in order to obtain estimates of the spectrum process
$\{S_{jt}=w_{jt}^2\}$, for each scale $j$. We form the vector
$y_t=(y_{1t},y_{2t})'$, where $y_{1t}$ is the log-return value of
GBP and $y_{2t}$ is the log-return value of EUR. For each series
$\{y_{1t}\}$ and $\{y_{2t}\}$, respectively, Figures \ref{fig2} and
\ref{fig3} show simulations of the posterior spectrum $\{S_{jt}\}$,
for scales 1 and 2. The smoothed estimates of these figures are
achieved by first computing the smoothed estimates using the Kalman
filter and then applying a standard Spline method (Green and
Silverman, 1994). We note that, for the data set considered in this
paper, the estimates of Figures \ref{fig2} and \ref{fig3} are less
smooth than those produced by the method of Nason {\it et al.}
(2000). However, a higher degree of smoothness in our estimates can
be achieved by considering small values of the variance
$\sigma_j^2$, which controls the smoothness of the shocks in the
random walk of the $w$'s.

\begin{thebibliography}{10}

\bibitem{Anderson}
Anderson, P.L. and Meerschaert, M.M. (2005) Parameter estimation for
periodically stationary time series. {\it Journal of Time Series
Analysis}, {\bf 26}, 489-518.

\bibitem{Dahlhaus}
Dahlhaus, R. (1997) Fitting time series models to nonstationary
processes. {\it Annals of Statistics}, {\bf 25}, 1-37.

\bibitem{Francq01}
Francq, C. and Zakoan, J.M. (2001)  Stationarity of multivariate
Markov-switching ARMA models. {\it Journal of Econometrics}, {\bf
102}, 339-364.

\bibitem{Fryz}
Fryzlewicz, P. (2005) Modelling and forecasting financial
log-returns as locally stationary wavelet processes. {\it Journal of
Applied Statistics}, {\bf 32}, 503-528.

\bibitem{Fryz2}
Fryzlewicz, P. and Nason, G.P. (2006) Haar-Fisz estimation of
evolutionary wavelet spectra. {\it Journal of the Royal Statistical
Society Series B}, {\bf 68}, 611-634.

\bibitem{Fryz3}
Fryzlewicz, P., Van Bellegem, S. and von Sachs, R. (2003)
Forecasting non-stationary time series by wavelet process modelling.
{\it Annals of the Institute of Statistical Mathematics}, {\bf 55},
737-764.

\bibitem{Green94}
Green, P.J. and Silverman, B.W. (1994) {\it Nonparametric Regression
and Generalized Linear Models: A Roughness Penalty Approach.}
Chapman and Hall.

\bibitem{Hallin}
Hallin, M. (1986) Nonstationary Q-dependent processes and
time-varying moving-average models - invertibility property and the
forecasting problem. {\it Advances in Applied Probability}, {\bf
18}, 170-210.

\bibitem{Mercurio}
Mercurio, D. and Spokoiny, V. (2004) Statistical inference for
time-inhomogeneous volatility models. {\it Annals of Statistics},
{\bf 32}, 577-602.

\bibitem{Nason}
Nason, G.P., von Sachs, R. and Kroisandt, G. (2000) Wavelet
processes and adaptive estimation of the evolutionary wavelet
spectrum. {\it Journal of the Royal Statistical Society Series B},
{\bf 62}, 271-292.

\bibitem{kostas}
Triantafyllopoulos, K. and Nason, G.P. (2007) A Bayesian analysis of
moving average processes with time-varying parameters. {\it
Computational Statistics and Data Analysis}, {\bf 52}, 1025--1046.

\bibitem{vBvS}
Van~Bellegem, S.\ and von~Sachs, R. (2008) Locally adaptive estimation
of evolutionary wavelet spectra. {\em Annals of Statistics}, (to appear).




\end{thebibliography}
\end{document}